\documentclass[10pt]{article}
\usepackage[explicit]{titlesec}
\usepackage{hyphenat}
\usepackage{ragged2e}
\RaggedRight

\usepackage[T1]{fontenc}
\usepackage{amsmath, amsthm}
\usepackage{graphicx}
\usepackage{soul}
\setul{.6pt}{.4pt}
\usepackage{geometry}
\usepackage{fancyhdr}
\usepackage[labelfont={footnotesize,bf} , textfont=footnotesize]{caption}
\makeatletter
\renewcommand\tagform@[1]{\maketag@@@ {\ignorespaces {\footnotesize{\textbf{Equation}}} #1.\unskip \@@italiccorr }}
\makeatother
\titleformat{\section}
  {\normalfont}{\thesection}{1em}{\MakeUppercase{\textbf{#1}}}
\titlespacing\section{0pt}{0pt}{-10pt}
\titleformat{\subsection}
  {\normalfont}{\thesubsection}{1em}{\textit{#1}}
\titlespacing\subsection{0pt}{0pt}{-8pt}

\makeatletter
\newcommand\sixteen{\@setfontsize\sixteen{17pt}{6}}
\renewcommand{\maketitle}{\bgroup\setlength{\parindent}{0pt}
\begin{flushleft}
\sixteen\bfseries \@title
\medskip
\end{flushleft}
\textit{\@author}
\egroup}
\makeatother
\usepackage[sort&compress]{natbib}
\makeatletter
\renewcommand\@biblabel[1]{\textbf{#1.}\hfill}
\makeatother

\bibpunct{}{}{,~}{s}{,}{,}
\usepackage{hyperref}
\title{A Naive Model of Covid-19 Spread From A Dry Cough}
\author{Cristian Ramirez Rodriguez\\ \medskip 
Saint Mary's College of California, Moraga, CA 
}

\usepackage{float}
\begin{document}

\vspace*{.01 in}
\maketitle
\vspace{.12 in}

\section*{abstract}
Health experts have suggested that social distancing measures are one of the most effective ways of preventing the spread of Covid-19. Research primarily focused on large Covid filled droplets suggested that these droplets can move further than regulated social distancing guidelines (2 meters apart) in the presence of wind. This project aims to model the paths of smaller Covid virions that last longer in the air to see if they also move beyond social distancing norms in the presence of wind. By numerically solving a 2-dimensional Langevin equation for 3000 particles and modeling wind as one-dimensional and steady, velocities were found that translated into particles' positions. With wind, infectious doses of virions appeared to travel farther and faster, increasing the risk of infection before dispersion. The two-dimensional model given implies that social distancing norms are reasonable in cases with no wind, yet not conservative enough when there is wind.  Keywords: Computational Physics, Covid-19, Langevin equation, Numerical modeling, Social distancing
\vspace{.12 in}


\section*{introduction}

In the Spring of 2020, as a response to the ongoing pandemic, numerous political entities, under the advice of medical professionals recommended social distancing measures of 2 meters to minimize the spread of Covid-19. Research soon came out that the virus could likely travel much farther than expected. A research duo from Cyprus, Dr. Talib Dbouk and Dr. Dimitris Drikakis, used a three-dimensional computational model to investigate how wind could affect the movement of Covid virions through air, finding that winds from 4 km/hr to 15 km/hr resulted in saliva droplets traveling up to 6 m. \cite{dbouk}  Later numerical papers appear to confirm those results, \cite{feng} and a researcher working with the Jakarta health department found that the range of virus spread can be predicted from the direction of wind. \cite{Rendana} Numerical research on the effect of wind on Covid spread has primarily concerned itself with saliva droplets, while research has been sparser on smaller Covid virions. This project aims to predict the behavior of dry Covid virions and whether, in the presence of wind, enough of those particles move beyond the 2m "safe" social distance zone to cause infection.

\section*{methods}
\textit{Mathematical Justification}

A single coronavirus virion has a diameter between 60-140 nanometers. \cite{bar} As such, it is within the acceptable range of sizes to use the Langevin equation for Brownian motion, \textbf{Equation 1} to describe its movements in the absence of wind. \cite{zwan}

\begin{equation}
\dfrac{d}{dt}v(t)= \dfrac{1}{m}[\xi(t)-6\pi\eta rv(t)] \label{equ1}
\end{equation}

Here $v(t)$ is the velocity of the Brownian particle, $\frac{d}{dt}v(t)$ is the the acceleration of the Brownian particle, $\eta$ is fluid viscosity, r is the radius of the particle, m is the mass of the particle and $\xi(t)$ (a random force) is a stochastic variable based on random collisions of air molecules with the particle. \textbf{Equation 1} is equivalent to Newton's second law where the acceleration of a particle is equal to the force of random collisions moving the particle minus the drag force operating in the opposite direction of the velocity all divided by the mass of the particle.

Solving this differential equation for velocity gives
\begin{equation}
v(t)=C_1e^{(-6\eta\pi rt)/m}+\frac{\int \xi(t) e^{(6\eta\pi rt)/m} \ dt}{m}e^{(-6\eta\pi rt)/m}
\label{equ2}
\end{equation}

At time $t=0$ there will be no random collisions and the exponential reduces to 1, $C_1=v_0$, the initial velocity of the particle.

The integral term is not trivial to solve analytically. \cite{Terl} The integral simplifies if $\xi(t)$ is approximated as a force with a constant magnitude going in a random direction (which can easily be simulated). The integral becomes a constant force  (which will be called $\kappa$) multiplied by the anti-derivative of an exponential ($\frac{m}{6\eta \pi r} e^{(6\eta \pi rt)/m}$). The exponentials and the masses cancel out, so \textbf{Equation 2} becomes \textbf{Equation 3}.

\begin{equation}
v(t)=v_0e^{(-6\eta \pi rt)/m}+\frac{\kappa}{6\eta \pi r} 
\label{equ3}
\end{equation}

What then is the force $\kappa$? By definition, it must be related to the average force per collision between the air and the particle. $F_{avg}$ can be quickly estimated applying an impulse-momentum approximation on a collision between a Covid virion and an air molecule. As the Covid virion is significantly larger and slower than an air molecule one can assume that an air particle completely bounces off with a speed near its original speed, in the opposite direction. As such
$F_{avg}\Delta t \approx m\Delta v$ and $\Delta v = 2v_{avg}$.

The average kinetic energy $<K>$ of an air molecule (where computationally the mass of an air particle $m_{air}$ is given as a molar mass, $M$ divided by Avogadro's number $N_A$) is given by \textbf{Equation 4}:
\begin{equation}
<K>=\frac{3}{2}kT = \frac{1}{2}m_{air}v_{avg}^2 = \frac{M}{2N_A}v_{avg}^2
\label{equ4}
\end{equation}

Rearranging for average speed in \textbf{Equation 4} and inputting the results into the mass-impulse relation leads to \textbf{Equation 5} where $m$ is the mass of the virion, $k$ is the Boltzmann constant, and $T$ is the temperature.

\begin{equation}
F_{avg} = \frac{2m\sqrt{\frac{3kT}{m_{air}}}}{\Delta t}
\label{equ5}
\end{equation}

Air molecules collide roughly 10 billion times per second, as such $\Delta t \approx 10^{-10}$ seconds.

$\kappa$ is the vector sum of the forces of all collisions. As every collision is approximated in this paper as having some average force, $\kappa$ can be considered to be an integer multiple, $N$, of $F_{avg}$ going in some arbitrary direction. 

\begin{equation}
\Vec{\kappa} = N\Vec{F}_{avg}
\label{equ6}
\end{equation}

One may notice in \textbf{Equation 3} that the effects of gravity are neglected as the downward terminal speed of small Covid-19 virions is negligible \cite{terminal} and can be shown to be negligible in \textbf{Equation 7} using Stokes Drag and approximating the particle as a sphere:

\begin{equation}
v_t\approx\frac{10^{-18}kg*9.8m/s^2}{70*10^{-9}m*1.81*10^{-5}kg/(m*s)}=0.000008 m/s
\label{equ7}
\end{equation}

\textit {A note on wind}

For the computational model, only steady winds will be considered. In this project, the random collisions of air are in two dimensions while steady wind can be considered a one dimensional constant force in the x-direction of the wind. If one wanted to consider more dimensions of wind one would have to render more dimensions of Covid particle flow, which, is possible, \cite{dbouk,feng} although beyond the scope of this naive model. Here, wind-speed is added to the x-component of a particle's velocity.

\textit{Computational model}

A single cough can produce 3000 Covid-19 virions.\cite{cough}
An assumption being made is that the initial velocity of each of virions is equal, and completely in the x-direction, this assumption is justifiable as it does not affect the final results. Looking at \textbf{Equation 3} and \textbf{Figure 1} in the \textbf{Results}, one sees that the only term with the initial velocity is attached to a rapidly decreasing exponential and does not affect motion.

Using Python3, the trajectories of individual particles after a cough were modelled and calculated for various scenarios using the following process:

\quad {1. A particle is assigned an initial x-velocity. Time, the random force and the particle’s y-velocity are set to zero.}

\quad {2. For each time step until it reaches a max time an angle is randomly chosen.} 

\quad {3. With angle $\theta$, to satisfy \textbf{Equation 6}, $v_x$ is defined as \textbf{Equation 3} with the term $\frac{\kappa}{6\pi\eta r}$ multiplied by $\cos(\theta)$.}

\quad{4. If there is wind $v_x=v_0e^{(-6\eta\pi rt)/m}+(\frac{\kappa}{6\pi\eta r})\cos{\theta}+v_{wind}$}

\quad {5. Similarly, $v_y=\frac{\kappa}{6\pi\eta r}\sin(\theta)$.}

\quad {6. The magnitude of the random force is defined as the previous random force plus or minus $\kappa$}.

\quad {7. $v_x$,$v_y$ and the current time-step are recorded in arrays}

\quad {8. Once the final time-step completes, x and y positions are recorded using scipy's built-in integration tool.}

\quad {9. The particle's final positions are recorded.}

The code used can viewed in the github repository: \href{https://github.com/cristianramirezrodriguez/naivemodelcovid}{cristianramirezrodriguez/naivemodelcovid}.

\section*{results}

Using steps 2 through 8 of the \textit{Computational model} I mapped out the rapid changes in velocity of one particle over 1 second. Due to randomness, the exact shape of the graphs varied, although some things remained constant no matter what.
As clearly seen in \textbf{Figure 1} and \textbf{Figure 2}, when there are no winds the horizontal and vertical velocities behave similarly. Note how the decreasing exponential of \textbf{Equation 3} rapidly reduces the effects of the initial velocity to zero while random motion defines the velocity from there on out.

\begin{figure}[H]
\centering
\includegraphics[width=0.75\textwidth]{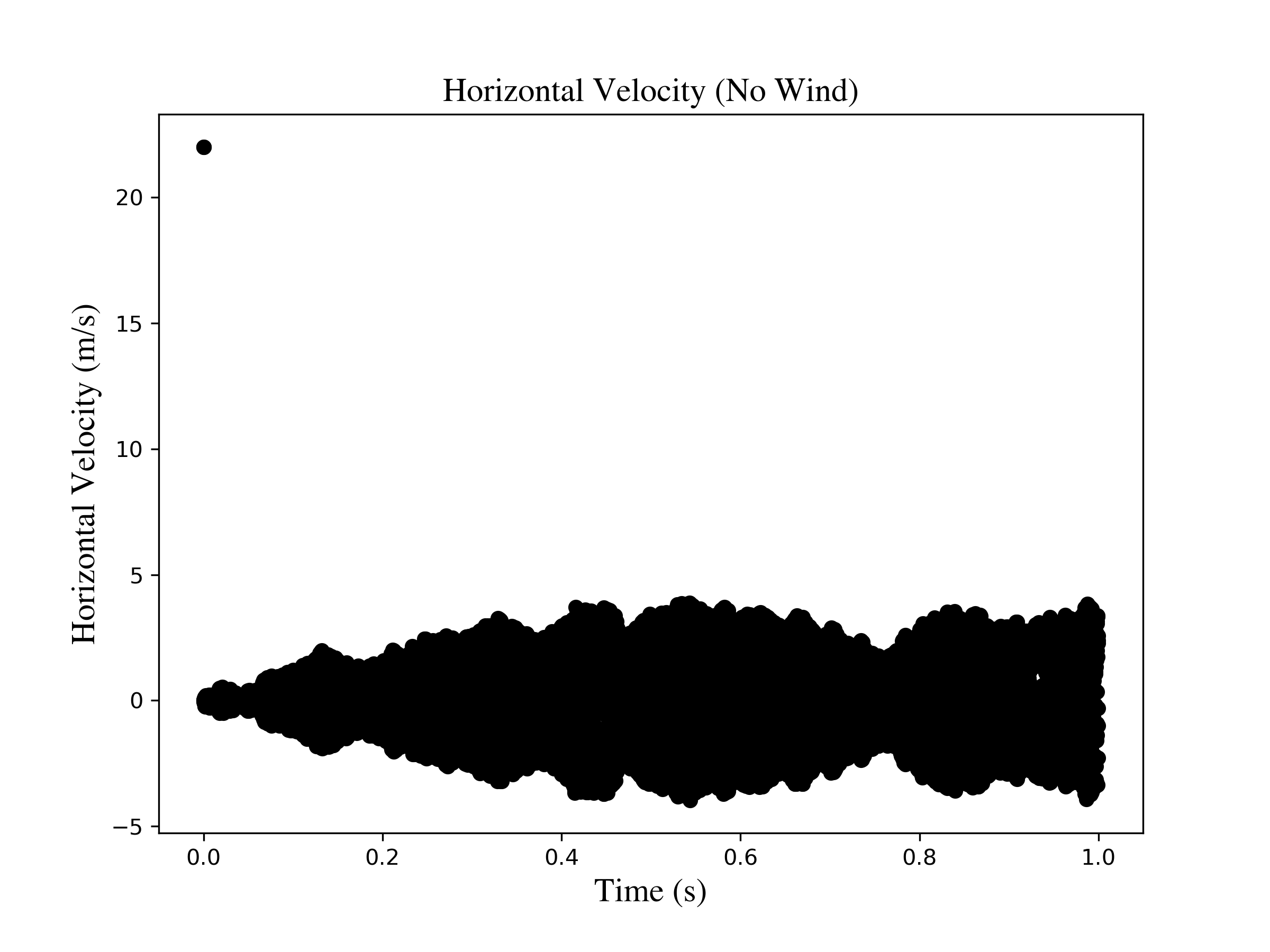}
\caption{The horizontal velocity of a single particle due to random collisions. When there is wind, the horizontal velocity graph appears similar, except centered at the wind velocity rather than zero.}
\label{figure 1}
\end{figure}

When there is wind, the horizontal velocity graph appears similar, except centered at the wind velocity rather than zero.

\begin{figure}[H]
\centering
\includegraphics[width=0.75\textwidth]{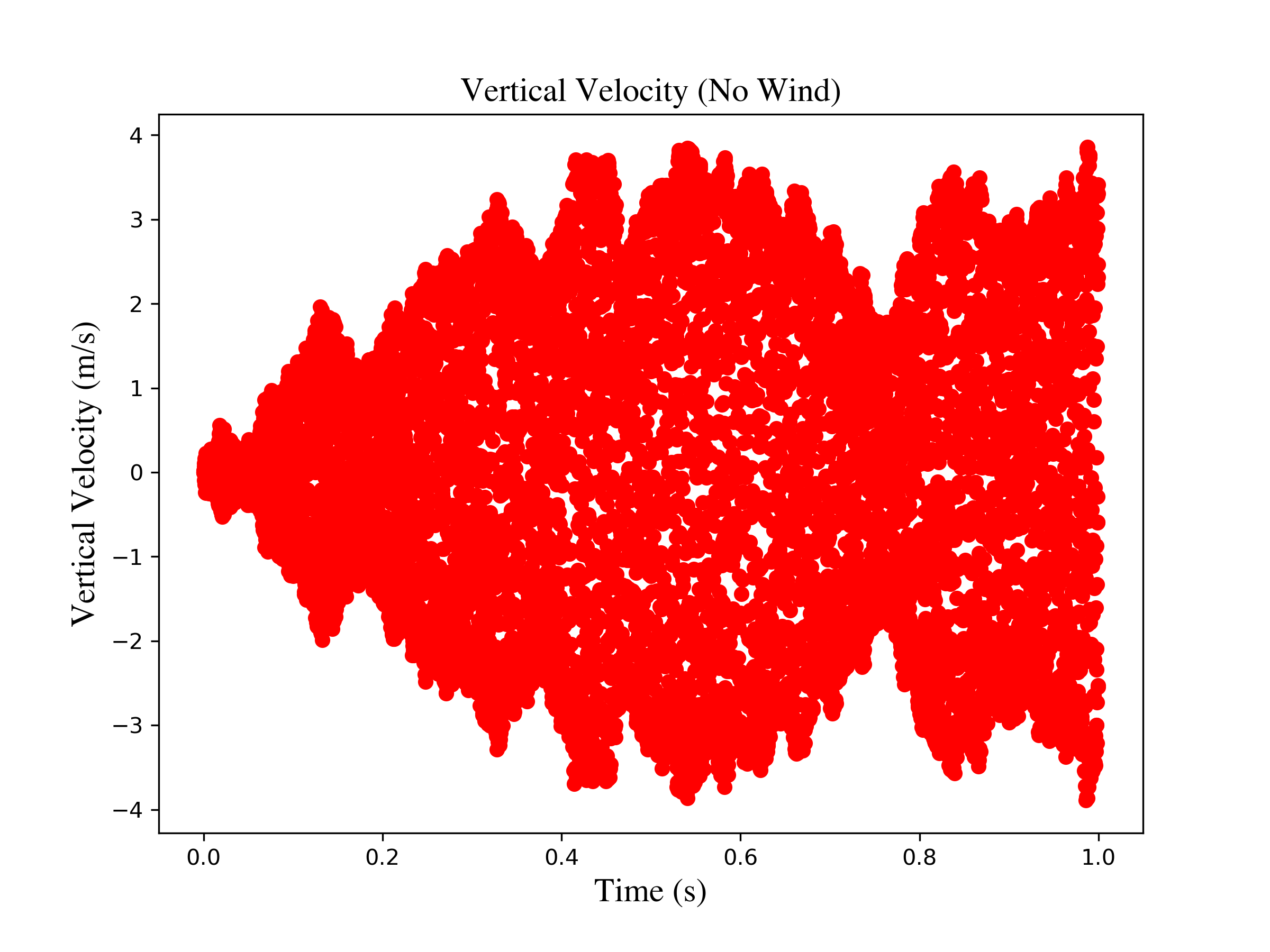}
\label{figure 2}
\caption{The vertical velocity of a single particle due to random collisions over the course of one second.}
\end{figure}

In both \textbf{Figure 1} and \textbf{Figure 2} the velocity appears to rapidly oscillate between -4 m/s and 4 m/s, with such a rapid shift, what if anything can be said about the position?

Running the previously mentioned nine step method shows that the radial distance the virus travelled without wind after 1 second is minimal as seen in \textbf{Figure 3}. Social distancing, according to this model, should prevent transmission in the absence of wind.

\begin{figure}[H]
\centering
\includegraphics[width=0.75\textwidth]{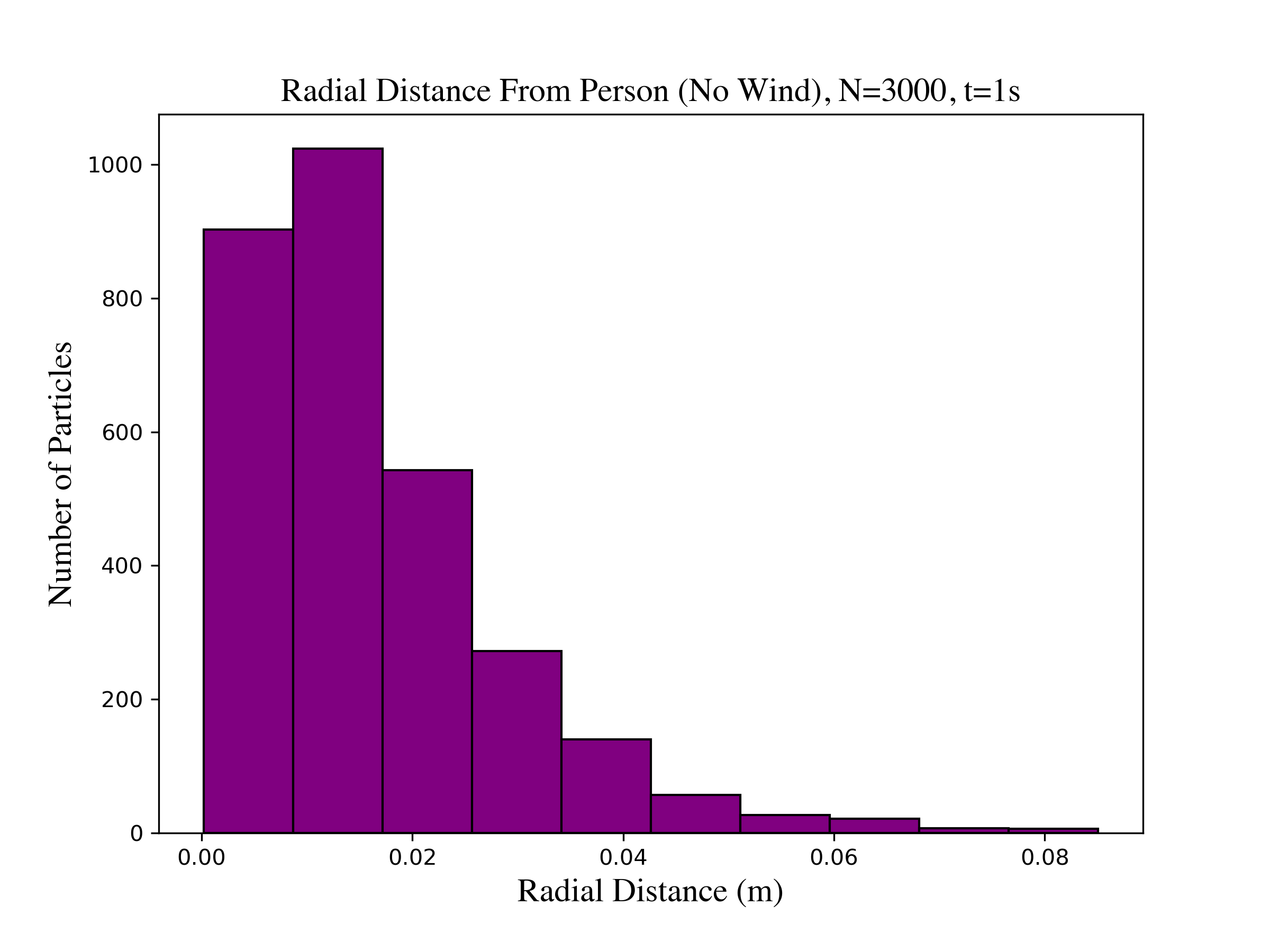}
\caption{One second after a cough the majority of covid virions unattached to saliva are closer than 0.02 meters from the source}\label{figure 3}
\end{figure}

Immunologist Dr. Bromage from the University of Massachusetts floated the number of 1000 virions as the minimum infectious dose. \cite{onek} Even if it is lower, such as 300, \cite{basu} it is comforting to predict that an infectious dose from a cough will not travel far without wind.

With wind, an infectious dose from a single cough travels roughly two meters in two seconds in 4km/hr winds as seen in \textbf{Figure 4}.

\begin{figure}[H]
\centering
\includegraphics[width=0.75\textwidth]{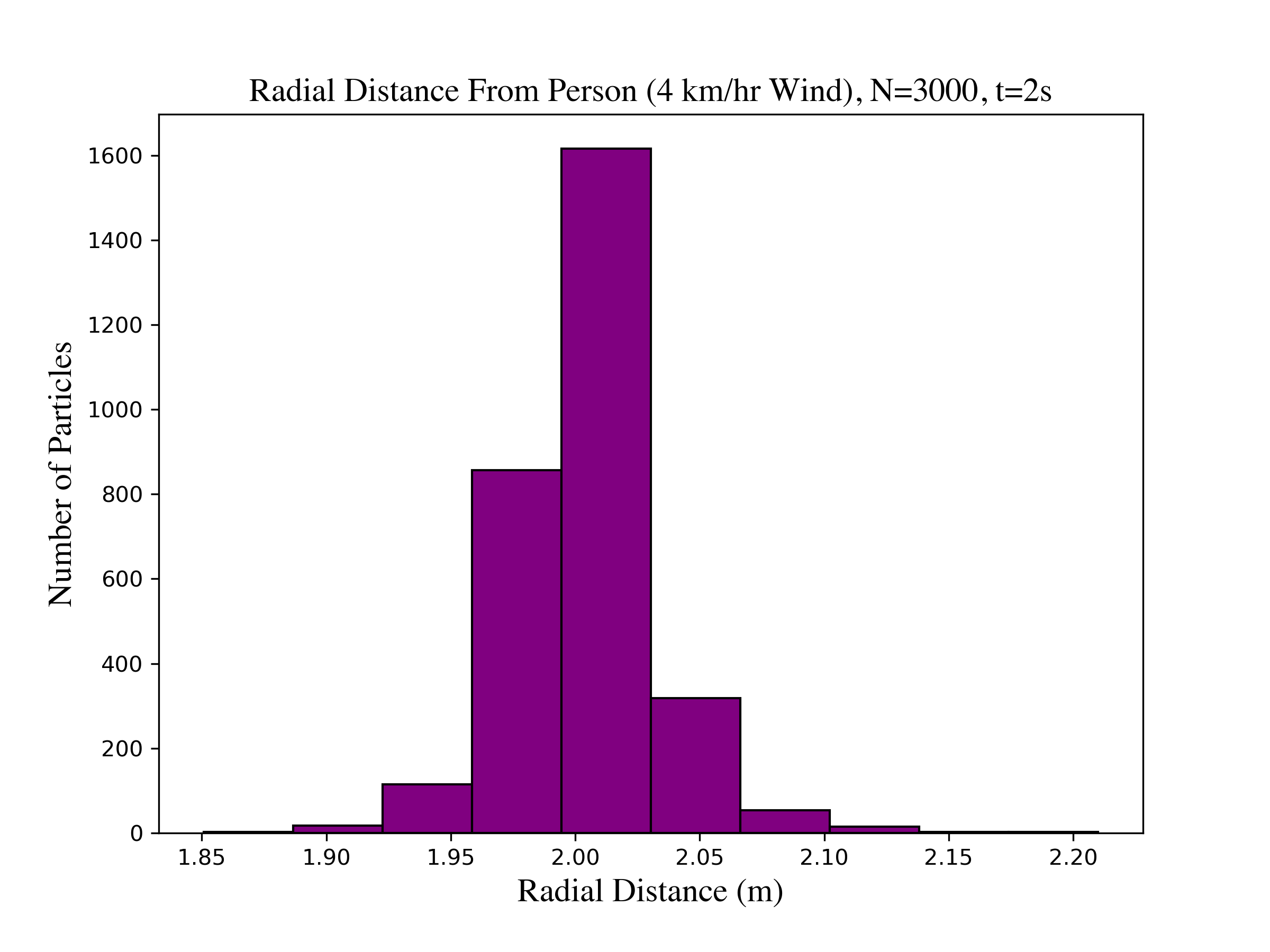}
\caption{In the presence of wind, virions travel greater distances.}\label{figure 4}
\end{figure}

That information, however, is not that important when considering spread. First, infection primarily occurs through the face, specifically the mouth and nose. \cite{fourth} So, the only virions in this model one should be worried about are those in the nose to chin area. Second, as time goes on virions disperse. Therefore, there should be a minimum safe time and distance where less than an infectious dose of Covid is in the danger region of an uninfected person’s face.

Using the assumption that both the infected and the potentially uninfected were of the same height, a while loop was used to keep moving forward through time until less than an infectious dose (1000 virions) exists in the danger area.  \textbf{Figure 5} and \textbf{Figure 6} show that without wind, roughly five seconds pass before an infectious dose leaves the danger zone. Note that this is a dispersion estimate only for a single cough producing 3000 virions; sneezing produces orders of magnitudes more, and talking can produce infectious doses as well. 

\begin{figure}[H]
\centering
\includegraphics[width=0.75\textwidth]{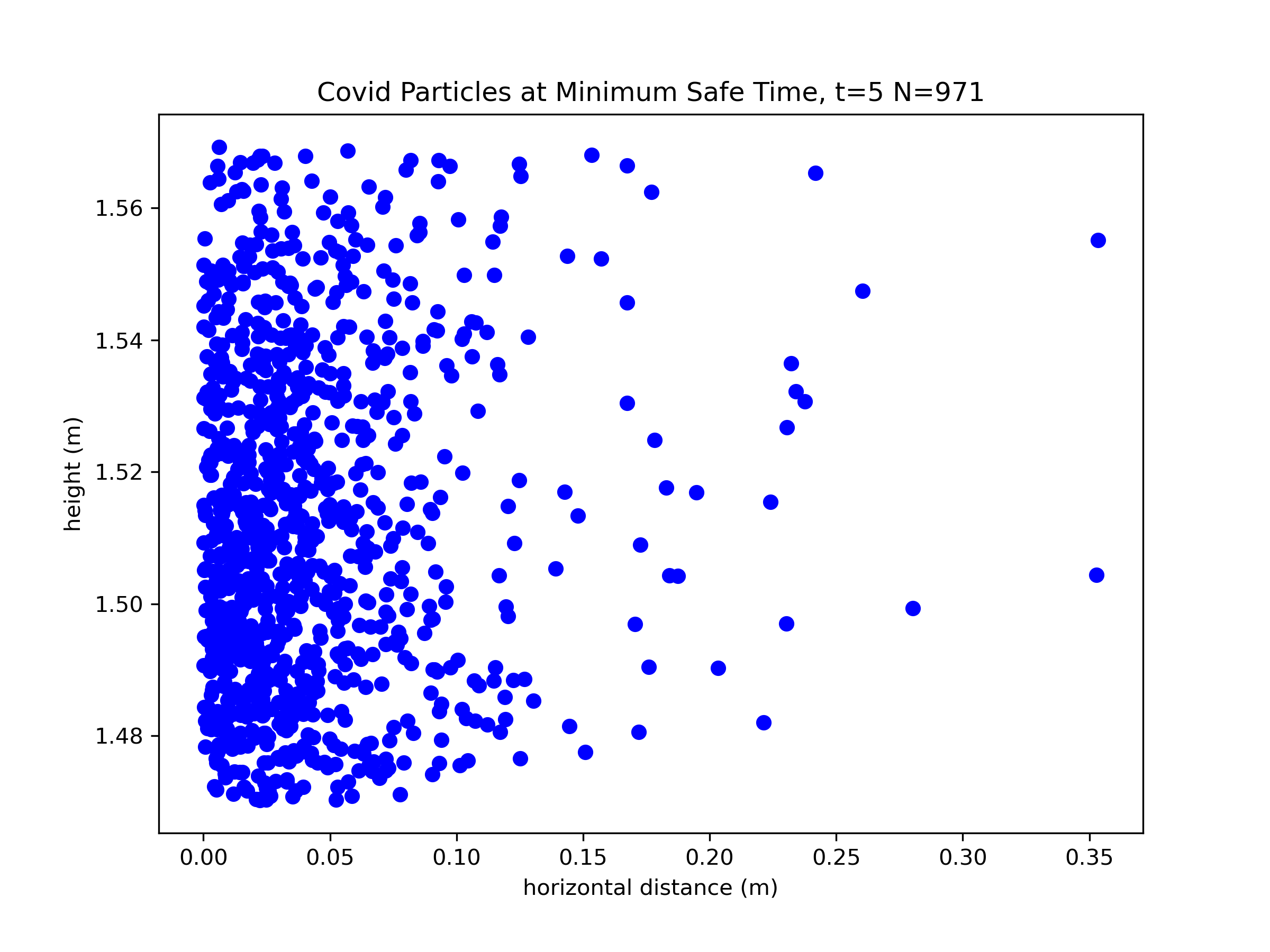}
\caption{In the simulation pictured, after 5 seconds, only 971 virions remain in the danger zone (at face level) and less than 0.4m away.}\label{figure 5}
\end{figure}

These figures have the majority of the virions within 0.40m which suggests that 2m social distancing is a safe distance without wind. 

\begin{figure}[H]
\centering
\includegraphics[width=0.75\textwidth]{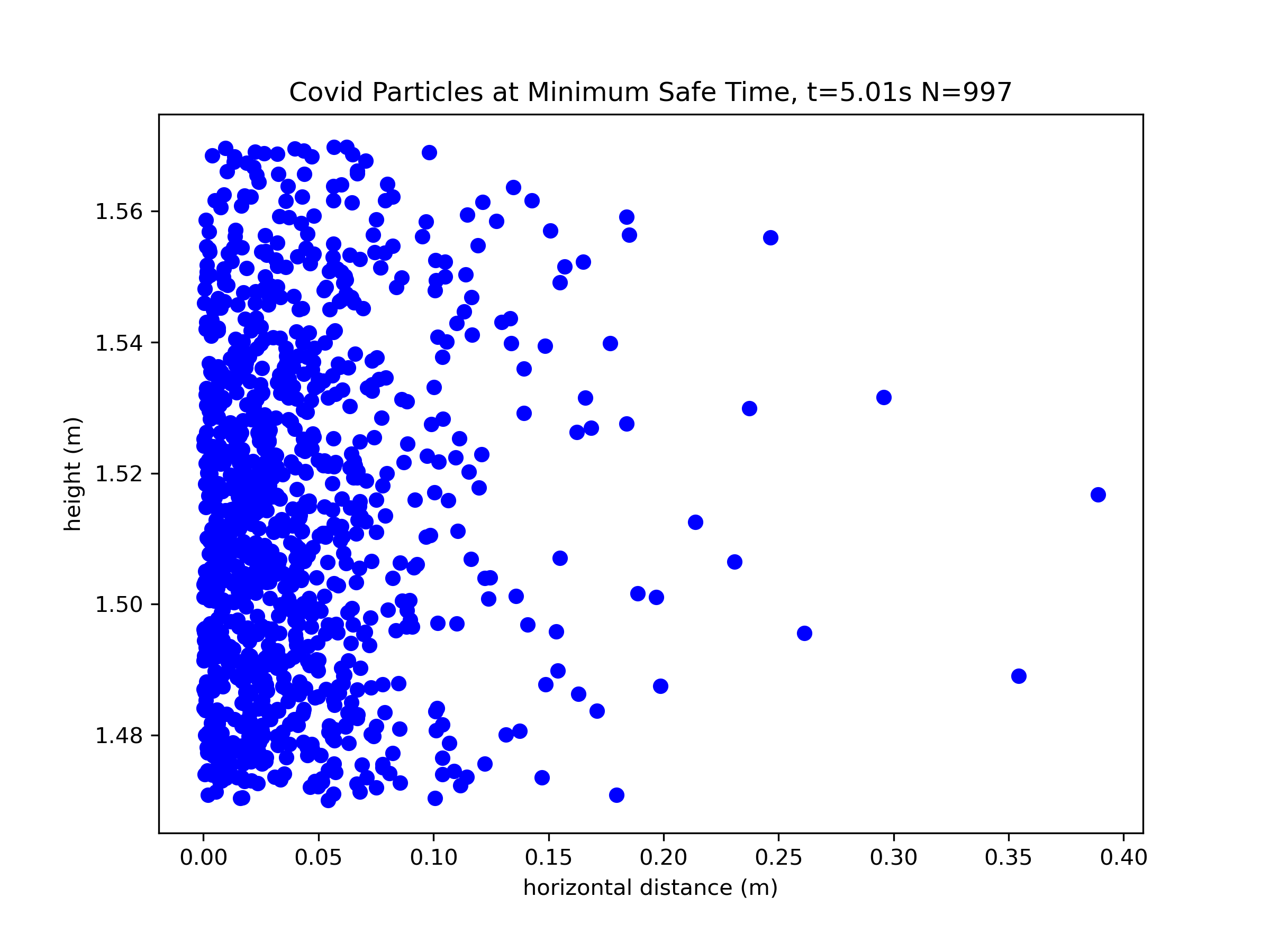}
\caption{In another simulation, it takes 5.01 seconds for there to be under 1000 virions in the danger zone.}\label{figure 6}
\end{figure}

With light, steady wind the virus goes farther and takes longer to dissolve in this model. This is shown clearly in \textbf{Figure 7} where roughly half of the original 3000 virions are still present 6 seconds later and six meters forward.

\begin{figure}[H]
\centering
\includegraphics[width=0.75\textwidth]{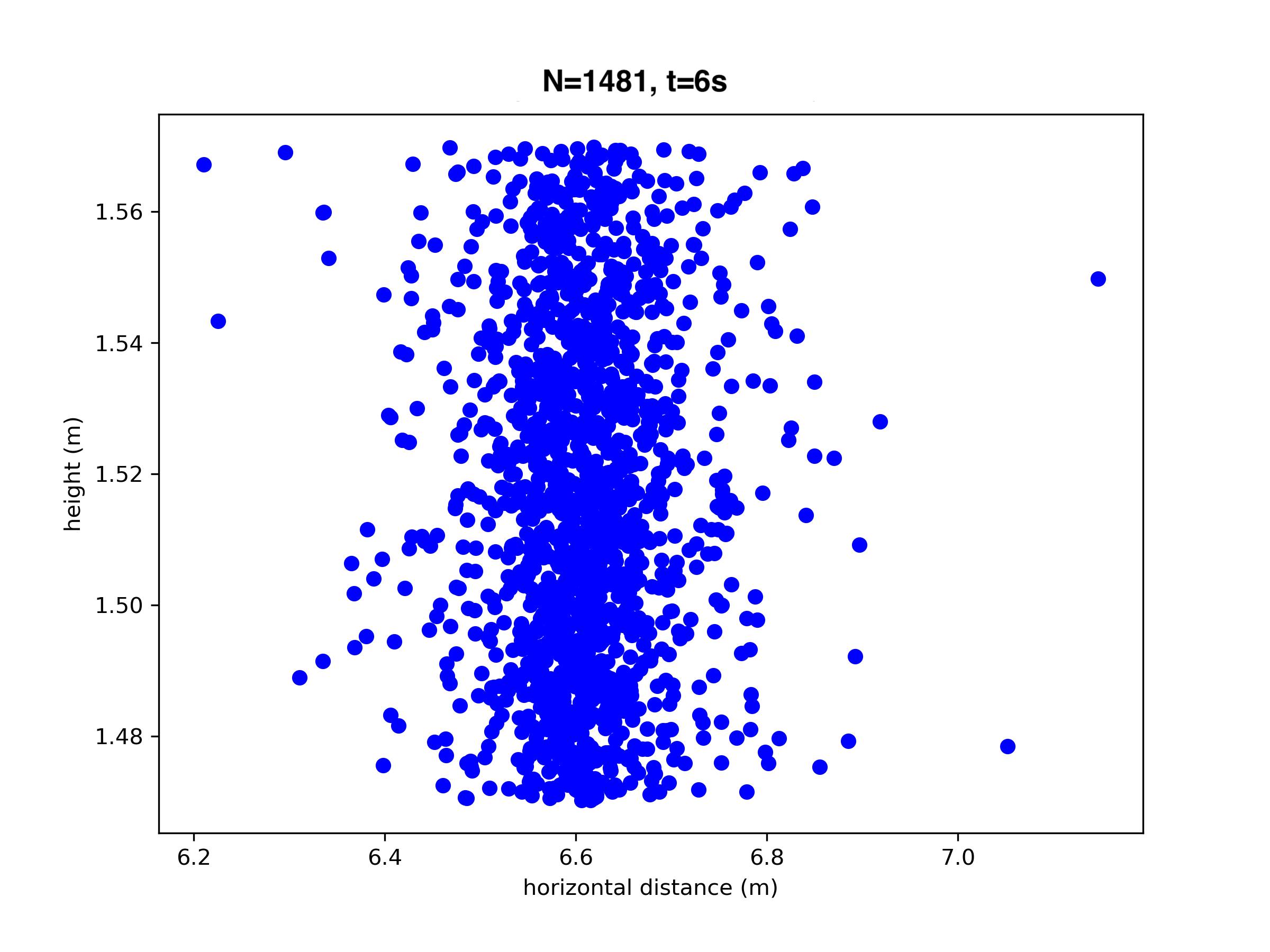}
\caption{A 2-dimensional person standing 2 meters away from the cougher would have received a face full of coronavirus in this simulation with wind.}\label{figure 7}
\end{figure}

\section*{discussion}
According to this model, in the presence of steady wind, infectious doses of Covid-19 can travel further and disperse slower, thereby increasing the risk of infection. This is in alignment with previous computational and physical research on larger droplets moving farther due to wind. \cite{dbouk, Rendana}

This naive model contains numerous limitations, one of them is that barring trade winds near the equator, few winds behave as one-dimensional.
Furthermore, living in a three-dimensional world there are more degrees of freedom and paths for particles to take. Those limitations could have resulted in an overestimation of the danger.
Another limitation was restricting this to a single dry cough to preserve computational power. Sneezes have a higher initial velocity, produce more virions, and produce more wet, larger particles that behave differently than the microscopic. With higher processing power this naive model could produce more accurate results.

\end{document}